# Design of a multifunctional sample probe for transport measurements and technological device characterizations

Mustafa ÖZTÜRK*, Numan AKDOĞAN
Gebze Technical University, Department of Physics, 41400 Gebze, Kocaeli, Turkey
*Correspondence: mozturk@gtu.edu.tr

**Abstract:** We describe a multifunctional sample probe for transport measurements, equipped with a two-axis goniometer providing computer-controlled 360° out-of-plane and manual 360° in-plane rotation. The multifunctional sample probe has been successfully implemented to a host dewar and electromagnet system. The developed probe is capable of performing transport measurements and device characterizations with high flexibility of controlling magnetic and electrical fields, angle, temperature and current/voltage parameters in a wide range. The design of the multifunctional sample probe allows easy connection of other external devices and easy installation into other dewar and magnet systems. A software interface based on NI Labview visual programming language has been developed to control the system. The setup has been tested in the hysteresis loop measurements of perpendicularly magnetized Co/Pt/CoO ultra-thin films with anomalous Hall effect (AHE) method. We have also performed temperature dependent exchange bias measurements of the samples. The developed probe revealed much better resolution in the case of perpendicularly magnetized ultra-thin Co/Pt/CoO films compared to a standard vibrating sample magnetometry system.

## 1. Introduction

Magnetotransport experiments are based on measurements of resistance, current or voltage characteristics of materials under the effect of magnetic field [1-6]. Modern developments in data processing and storage technologies involve further progress in magnetotransport methods. Furthermore, many trend topics like skyrmions [7, 8], magnetocaloric effect [9-12], spin Hall effect [10, 13, 14] and planar Hall effect for biosensor applications [15] are also studied by transport measurements. More energy efficient, much faster, smaller and non-volatile devices are strongly desirable for many technological applications. On the other hand, in order to develop these new generation devices practical, fast and flexible experimental instrumentation is required.

Furthermore, the type of the material under investigation and the measurement conditions are important factors for planning a transport experiment. In many cases, it is required to rotate the sample at a fixed temperature without taking the sample out and stopping the experiment. Besides the magnetic field, electrical field is also widely used to characterize the technological devices and to manipulate the magnetization of ultra-thin films and bulk materials [16, 17]. For this reason, a possibility to apply the electrical field is very desirable option of a transport setup. Depending on the material type and other needs, it is highly practical to have a wide range temperature, magnetic field, electrical field, angle, current and voltage parameters in a single setup.

In this study, we report the design and construction of a multifunctional sample probe with a rotatable sample holder. We successfully tested this probe in combination with host dewar and magnet system. The developed setup allows controlling of electrical and magnetic fields, sample angle, temperature and current/voltage values in a wide range for many kinds of transport measurements using a single system. In order to test the system anomalous Hall effect (AHE) measurements for perpendicularly magnetized ultra-thin films have been carried out and hysteresis loops for perpendicularly magnetized Co(4 Å) /Pt /CoO and Co(5 Å) /Pt/ CoO test samples have been obtained. The temperature dependence of exchange bias effect in the samples has also been investigated by the AHE measurements at different temperatures.

## 2. Instrument

In order to carry out various transport measurements at different geometries in a single setup, a careful design of a sample holder is required. Therefore, we designed a rotatable sample holder with a possibility of computer-controlled 360° out-of-plane and manual 360° in-plane sample rotation in vacuum conditions for a wide temperature range.

Figure 1 shows the schematic drawings with the dimensions of multifunctional sample probe. The total length of the sample probe and diameter of the sample holder are 878.4 mm and 25.5 mm, respectively. The vertical and horizontal position of the sample is set to correspond to the center of the superconducting magnet within the host dewar. The computer controlled out-of-plane rotation of the sample holder is provided by a stepper motor located on top end of the sample probe, with a precision of 0.125°. Figure 2 (left) shows a picture of the sample holder attached to the bottom end of the probe. The sample holder includes a printed circuit board (PCB) sitting on the axis of a gear wheel. The samples with the sizes up to 12 mm x 12 mm can be mounted on a pluggable male PCB. The sample probe and sample holder are made from nonmagnetic stainless steel and brass material, respectively.

The multifunctional sample probe includes a vacuum-tight connector on top and it is connected to a sample bridge via an insulated cable. The sample bridge is designed as a cable distribution and connection box. It consists of a 12 contact holes for the connections between the sample holder and all other external devices as shown in Figure 2 (right). Four of the connection holes located on the sample bridge are reserved for the cernox thermometer and 8 holes are used for the measurement devices. The sample bridge can also be used as a test measurement tool. It has a female PCB which allows testing the connections of the sample before the measurements.

A general scheme of the whole transport measurement setup consisting of the multifunctional sample probe, sample bridge, external devices, the host dewar and electromagnet system is shown in Figure 3. With this design, installation and de-installation of the sample probe can be done within seconds. In the test measurements, QD PPMS9T has been used as a host dewar, which provides the magnetic fields up to 9 T within the temperature range of 2 K to 400 K. A current source is combined with a voltmeter for the resistance measurements of low resistant materials. For the samples with higher resistivity (semiconductors or insulators) a high resistance meter is employed in order to apply high voltages and measure low currents. It also serves as a voltage source for electrical field manipulated AHE measurements. Data acquisition is realized by the general-purpose interface bus (GPIB) connected to the external devices. The magnetic field, vacuum and temperature control of the whole setup is managed via control unit of the host system. In addition to the temperature sensors of the host system, the sample probe includes a calibrated cernox temperature sensor attached to the sample holder. This sensor provides more accurate setting and measurement of temperature in the location of a sample.

Since the system has many variables (system and sample temperatures, sample angle, pressure control, magnetic and electrical fields, current and voltage), practical control of the devices and data acquisition needs in fully automated and computer controlled interface. For this reason, we developed a user-friendly software based on NI Labview. Figure 4 demonstrates the user interface of the software. Real-time information and the data from all devices of the setup are always monitored by the user interface. The software allows users to simultaneously plot the data of present and previous measurements.

### 3. Test Measurements

Since AHE is one of the most convenient method to test and calibrate all parameters of the multifunctional sample probe, we have carried out AHE experiments with perpendicularly magnetized Pt(5 Å) /Co(5 Å) /Pt(5 Å) /CoO(10 nm) /Pt(3 nm) (sample A) and Pt(5 Å) /Co(4 Å) /Pt(5 Å) /CoO(10 nm) /Pt(3 nm) (sample B) ultra-thin films grown on fused silica substrates. For the observation of AHE contribution in thin films, ferromagnetic layer should be perpendicularly magnetized or should have a perpendicular projection of magnetization. Hall resistivity ($\rho_H$) of these kind of thin films is the sum of two contributions. The first one is an ordinary Hall effect ($\rho_{HE}$) and the second one is due to anomalous Hall effect ($\rho_{AHE}$) [6]. In ferromagnetic materials, since AHE is often much larger than the ordinary Hall effect, the Hall resistivity measurements reflect the magnetization changes and results in a hysteresis loop [6]. This can be expressed as

$$\rho_H = \rho_{HE} + \rho_{AHE} = R_0 H + R_A \mu_0 M \quad (1)$$

where $R_0$, $H$, $R_A$, $\mu_0$ and $M$ stand for the Hall coefficient, applied magnetic field, anomalous Hall coefficient, permeability of vacuum and the material magnetization respectively.

Figure 5 shows the hysteresis loop measurements of sample A obtained by both VSM and AHE methods. It is clearly seen from the figure that the hysteresis curve measured by VSM has lower signal-to-noise ratio (SNR) and the diamagnetic contribution of the substrate is very dominant. By the way, the measurement with AHE has much better SNR and the square shape of the loop is more obvious. This is due to inherently better sensitivity of AHE method for ultra-thin ferromagnetic materials. Since the methods like VSM [18] and superconducting quantum interference device (SQUID) [19] are bulk measurement techniques, ferromagnetic properties of ultra-thin films are usually measured against a strong background due to diamagnetic or paramagnetic contributions from the substrate, holder or other layers.

Figure 6 presents the temperature dependent AHE hysteresis loop measurements of sample B. The loops are measured at temperatures of 250 K, 275 K and 300 K. Since the sample has both ferromagnetic (FM) Co and antiferromagnetic (AF) CoO layers, a shift in the hysteresis loop of the FM layer was observed at the temperatures below the Nèel temperature of CoO (~290 K). This shift is called as exchange bias (EB) effect [20, 21], which has great importance for data storage and magnetic sensor technologies. These results prove the capability and sensitivity of the developed probe for magneto-transport measurements of ultra-thin films.

### 4. Conclusion

In summary, a design of a multipurpose sample probe with a rotatable sample holder and its successful integration into a host electromagnet system have been presented. The setup allows to perform magnetotransport measurements at various magnetic and electrical fields, angle, temperature and current/voltages which are controllable in a wide range. The setup is flexible in configuration and allows adding/removing any external device according to the user demands. The multifunctional sample probe can also be used with other dewar and magnet systems for many other possible applications. We also developed user-friendly software which provides control of the whole system and real-time data acquisition. We have successfully tested the system in AHE measurements of ultra-thin films. Square-like hysteresis loops and EB effect have been observed in the perpendicularly magnetized Co (4 Å) /Pt /CoO and Co (5 Å) /Pt /CoO ultra-thin films with a very good sensitivity.

**Acknowledgments**

We would like to acknowledge Osman Öztürk and Melek Türksoy Öcal from Gebze Technical University for preparation of the ultra-thin ferromagnetic films. We would like to acknowledge Bulat Rami, Mustafa Özdemir and Ali Cemil Başaran for fruitful discussions. We also would like to thank Münir Dede, Serhat Çelik, Örgür Karcı and Ahmet Oral from NanoMagnetics Instruments for manufacturing of the sample probe and the sample bridge according to our preliminary design and requirements. We also gratefully acknowledge Joachim Speck, Jörg Tobish, Oleg Ignatchik, Stefan Riesner and Neil Dilley from LOT-Quantum Design for their continuous technical support related to the host PPMS system. This work was partially supported by TÜBİTAK (The Scientific and Technological Research Council of Turkey) Grant No 112T857 and by DPT (State Planning Organization of Turkey) through the project number 2009K120730.

**References**

[1] Thomson, W. *P R Soc London* **1856**, 8, 546-550.

[2] Baibich, M. N.; Broto, J. M.; Fert, A.; Nguyen Van Dau, F.; Petroff, F.; Etienne, P.; Creuzet, G.; Friederich, A.; Chazelas, J. *Phys Rev Lett* **1988**, 61, 2472-2475.

[3] Binasch, G.; Grunberg, P.; Saurenbach, F.; Zinn, W. *Phys Rev B* **1989**, 39, 4828-4830.

[4] Moodera, J. S.; Kinder, L. R.; Wong, T. M.; Meservey, R. *Phys Rev Lett* **1995**, 74, 3273-3276.

[5] Hall, E. H. *Am J Math* **1879**, 2, 287-292.

[6] Nagaosa, N.; Sinova, J.; Onoda, S.; MacDonald, A. H.; Ong, N. P. *Rev Mod Phys* **2010**, 82, 1539-1592.

[7] Schulz, T.; Ritz, R.; Bauer, A.; Halder, M.; Wagner, M.; Franz, C.; Pfleiderer, C.; Everschor, K.; Garst, M.; Rosch, A. *Nat Phys* **2012**, 8, 301-304.

[8] Yu, X. Z.; Kanazawa, N.; Zhang, W. Z.; Nagai, T.; Hara, T.; Kimoto, K.; Matsui, Y.; Onose, Y.; Tokura, Y. *Nat Comms* **2012**, 3, 988.

[9] Kim, H.; Huse, D. A. *Phys Rev A* **2012**, 86, 053607.

[10] Uchida, K.; Takahashi, S.; Harii, K.; Ieda, J.; Koshibae, W.; Ando, K.; Maekawa, S.; Saitoh, E. *Nature* **2008**, 455, 778-781.

[11] Meier, D.; Reinhardt, D.; van Straaten, M.; Klewe, C.; Althammer, M.; Schreier, M.; Goennenwein, S. T. B.; Gupta, A.; Schmid, M.; Back, C. H., et al. *Nat Comms* **2015**, 6, 1-7.

[12] Flipse, J.; Bakker, F. L.; Slachter, A.; Dejene, F. K.; van Wees, B. J. *Nat Nanotechnol* **2012**, 7, 166-168.

[13] Hirsch, J. E. *Phys Rev Lett* **1999**, 83, 1834-1837.

[14] Okamoto, N.; Kurebayashi, H.; Trypiniotis, T.; Farrer, I.; Ritchie, D. A.; Saitoh, E.; Sinova, J.; Mašek, J.; Jungwirth, T.; Barnes, C. H. W. *Nat Mater* **2014**, 13, 932-937.

[15] Hung, T. Q.; Terki, F.; Kamara, S.; Kim, K.; Charar, S.; Kim, C. *J Appl Phys* **2015**, 117, 154505.

[16] Chiba, D.; Fukami, S.; Shimamura, K.; Ishiwata, N.; Kobayashi, K.; Ono, T. *Nat Mater* **2011**, 10, 853-856.

[17] Chiba, D.; Ono, T. *J Phys D Appl Phys* **2013**, 46, 213001.

[18] Foner, S. *Rev Sci Instrum* **1959**, 30, 548-557.

[19] Jaklevic, R. C.; Lambe, J.; Silver, A. H.; Mercereau, J. E. *Phys Rev Lett* **1964**, 12, 159-160.

[20] Akdoğan, N.; Yağmur, A.; Öztürk, M.; Demirci, E.; Öztürk, O.; Erkovan, M. *J Magn Magn Mater* **2015**, 373, 120-123.

[21] Demirci, E.; Öztürk, M.; Sınır, E.; Ulucan, U.; Akdoğan, N.; Öztürk, O.; Erkovan, M. *Thin Solid Films* **2014**, 550, 595-601.

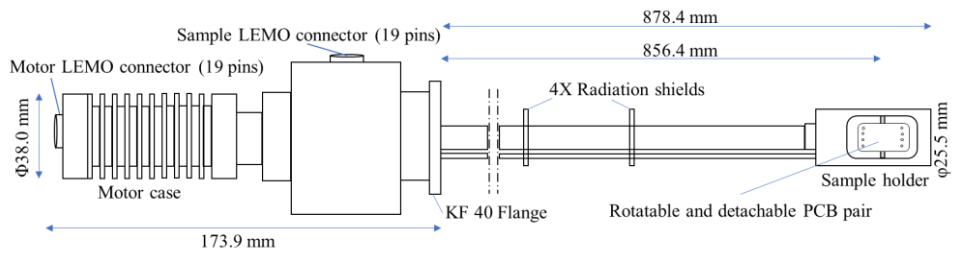

**Figure 1.** Schematic drawing of the multifunctional sample probe.

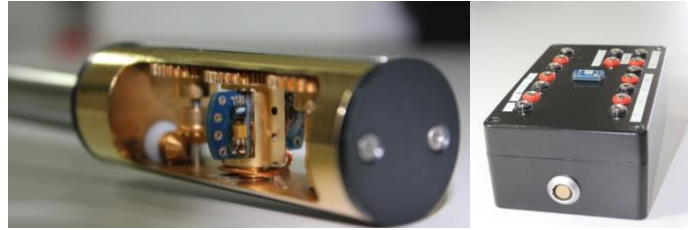

**Figure 2.** Pictures of the sample holder (left) and sample bridge (right).

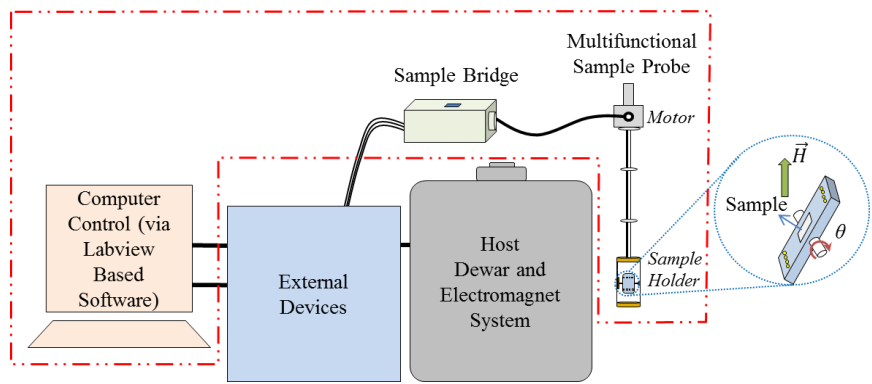

**Figure 3.** General view of the setup including multifunctional sample probe, host system and external devices.

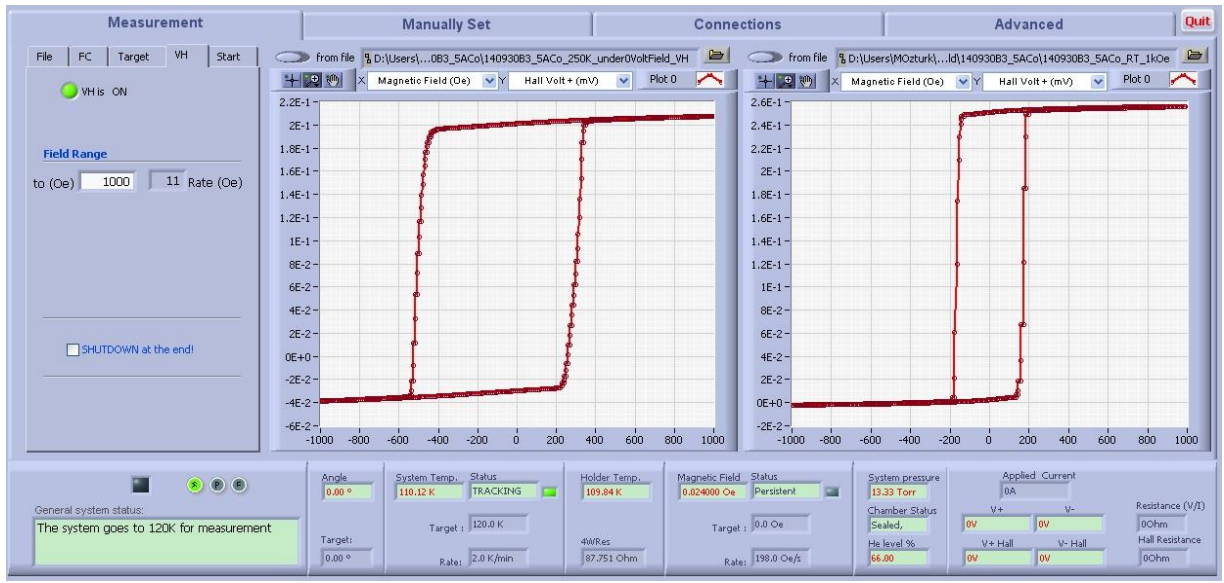

**Figure 4.** The main user panel of the software which allows controlling and reading all parameters of the measurement setup.

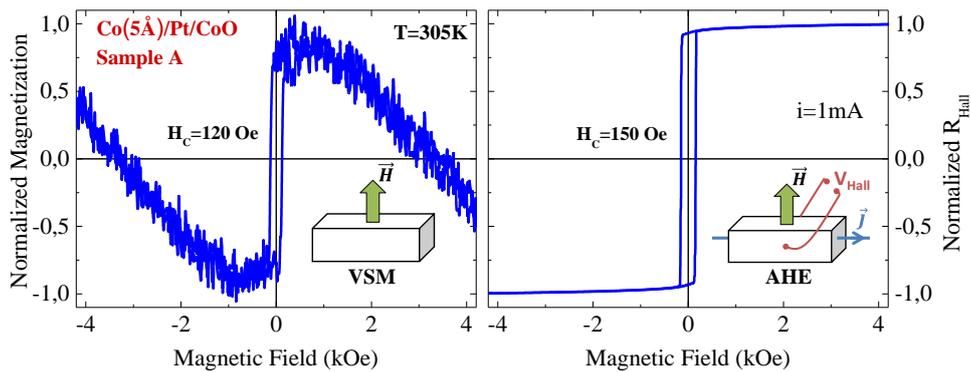

**Figure 5.** Perpendicular magnetization measurements of Co(5 Å)/Pt/CoO ultra-thin film (sample A) obtained by VSM (left) and AHE (right) techniques.

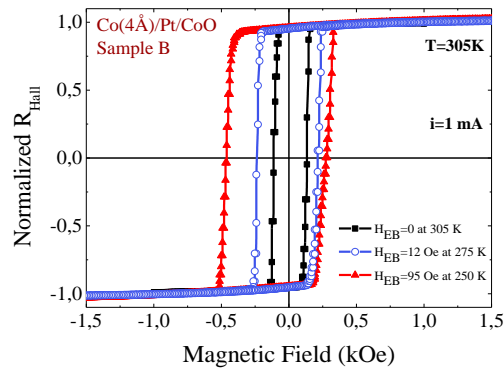

**Figure 6.** Exchange bias effect measurements of Co (4 Å) /Pt /CoO ultra-thin film were carried out by using AHE method at different temperatures.